\documentclass[twocolumn, letter]{jpsj3}
\usepackage{txfonts}
\usepackage{graphicx}% Include figure files
\usepackage{dcolumn}% Align table columns on decimal point
\usepackage{bm} % bold math
\usepackage{color}
\definecolor{green}{rgb}{0.0, 0.5, 0.0} 

\usepackage{ulem}

\title{Coulomb Frustrated Phase Separation in Quasi-Two-Dimensional Organic Conductors
\\ 
on the Verge of Charge Ordering}

\author{Kazuyoshi \textsc{Yoshimi}$^1, 2$ and Hideaki \textsc{Maebashi}$^3$}
\inst{$^1$Department of Physics, University of Tokyo, Tokyo 113-8656\\ $^2$Nanosystem Research Institute "RICS,"
National Institute of Advanced Industrial Science and Technology, Ibaraki 305-8568, Japan\\
$^3$Institute for Solid State Physics, University of Tokyo, Kashiwanoha, Kashiwa 277-8581, Japan}

\abst{
We study a $3/4$-filled two-dimensional extended Hubbard model under the fluctuation-exchange approximation. 
We find that this model undergoes phase separation in a region of nonzero temperature, where the quantum 
critical phenomenon of charge ordering dominates. 
By considering the long-range Coulomb interaction that frustrates this phase separation, 
we present a mechanism for generating a glassy state in a quasi-two-dimensional organic conductor 
on the verge of charge ordering.
}
\kword{organic conductors, charge ordering, phase separation, glassiness, quantum critical phenomena}

\def\runauthor{K. \textsc{Yoshimi} and H. \textsc{Maebashi}}

\begin{document}
\maketitle
Organic conductors are providing a fruitful stage for studying exchange-correlation effects between electrons on a crystal lattice. 
One of the reasons is because they are mutually interacting electron systems with both intra- and intersite Coulomb repulsions. 
The former characteristic gives rise to correlation effects and the latter to exchange-correlation effects, and both play essential 
roles in the interesting physics of these systems~\cite{Seo04}. In addition, they are very pure systems so their observable 
properties are not caused by impurities, which is an aspect that facilitates the extraction of pure effects that are due to mutual 
interactions between electrons~\cite{Singleton02,Sasaki}.  

Among the various phenomena characteristic of organic conductors, Wigner-type charge ordering (CO) is particularly interesting 
because it originates from intersite Coulomb repulsion~\cite{Seo04}. Surprisingly, recent $^{13}$C-NMR measurements of 
$\theta$-ET$_2$MM'(SCN)$_4$ (ET = BEDT-TTF, M = Rb, Cs, M' = Zn, Co) have revealed the existence of a ``glassy" state 
in quasi-two-dimensional (quasi-2D) organic conductors on the verge of CO, which is a spatially inhomogeneous state 
characterized by a slow charge relaxation with non-Arrhenius behavior~\cite{Miyagawa00}. X-ray diffraction measurements have indicated that 
this glassy state has an ``intermediate-range order", namely, the correlation length develops well but remains finite in the range of tens of nanometers~\cite{Watanabe99}, which is inseparably connected to giant nonlinear conduction, which may act as  a thyristor inverter in organic electronics~\cite{Sawano05}.

Because of the pureness of organic conductors, we assume that their glassy state does not come from quenched disorder 
but from the mutual interaction between electrons. Although it is not well clarified,
some efforts to identify this glassy state have been made on the basis of the 2D extended Hubbard model (EHM) with 
not only intrasite Coulomb repulsion but also intersite Coulomb repulsion~\cite{Merino05}. 

Apart from these efforts, there exists a theory of glassiness that does not rely on the presence of quenched disorder 
but is based on phase separation (PS) that is frustrated by the long-range Coulomb interaction~\cite{Schmalian00, Westfahl01}.
Examples of mutually interacting electron systems that strongly tend toward PS include the $t$-$J$ model~\cite{Emery90}, 
double-exchange model~\cite{Yunoki98}, and low-density electron gases~\cite{Vosko80}.
Theories involving Coulomb frustrated PS~\cite{Ortix08} have been applied to several classes of materials such as 
high-$T_{\rm c}$ cuprates~\cite{Emery93}, colossal magnetoresistive manganites~\cite{Yunoki99}, and supercritical-fluid 
alkali metals~\cite{Maebashi09}, but have yet to be applied to $\theta$-ET organic conductors. 

In this Letter, we report our theoretical results for the charge compressibility $\kappa$. These results are obtained by 
applying the fluctuation-exchange (FLEX) approximation to a simple model of $\theta$-ET organic conductors, namely, 
a $3/4$-filled 2D EHM with both on-site and nearest-neighbor Coulomb repulsion. We show that, in a quantum critical 
regime dominated by quasiclassical CO fluctuations, $\kappa^{-1}$ decreases with decreasing temperature and eventually 
becomes {\it negative}, which indicates that this model has a strong tendency for PS at {\it nonzero} temperature. 
By using our numerical results for $\kappa$, we estimate the correlation length $\xi$ and the period $l_{\rm m}$  
of the electron density modulation induced by the long-range Coulomb interaction that frustrates the PS. 
For $\theta$-ET organic conductors on the verge of CO, 
we find that the ratio $\xi /l_{\rm m}$ can be larger than the critical value close to $2$, indicating the emergence of 
a glassy state, i.e., a spatially inhomogeneous state characterized by an extremely slow relaxation and an intermediate length scale.

% model + approximation
Our EHM Hamiltonian is defined on a 2D square lattice 
with a lattice constant $a_{\rm L}$ as
\begin{equation}
H = t \! \sum_{\langle i,j \rangle, \sigma} 
( c_{i\sigma}^{\dagger} c_{j\sigma}  +  {\rm H.c.} )
+ 
U \! \sum_{i} {\tilde n}_{i\uparrow} {\tilde n}_{i\downarrow} + V \! \sum_{\langle i,j \rangle} 
{\tilde n}_i {\tilde n}_j 
\label{eq:ham}.
\end{equation}
Here, $t$ is the hopping integral between the nearest-neighbor sites, which are denoted by $\langle i, j \rangle$, 
$c_{i \sigma}^{\dag}$ ($c_{i \sigma}$) is the creation (annihilation) operator of 
an electron on site $i$ with a spin $\sigma$ $=$ $\uparrow$ or $\downarrow$, 
${\tilde n}_{i \sigma} = c_{i\sigma}^{\dag}c_{i\sigma} - n/2$ with $n$ being the mean electron number per site, 
${\tilde n}_{i} ={\tilde n}_{i\uparrow} + {\tilde n}_{i\downarrow}$, 
and $U$ and $V$ are the on-site and nearest-neighbor Coulomb 
repulsions, respectively. 
We use $a_{\rm L}=1$ and $t=1$ unless otherwise specified.

We calculate the charge compressibility $\kappa \equiv n^{-2} (\partial \mu / \partial n)^{-1}$ 
for the Hamiltonian eq.~(\ref{eq:ham}) at the temperature $T$. Here, the chemical potential $\mu \equiv \mu(n)$ 
is obtained as a function of $n$ using
\begin{equation}
n=2\int_k G(k){\rm e}^{{\rm i}\omega_k \eta},
\label{eq:n}
\end{equation}
where $\eta$ is a positive infinitesimal and $\int_{k}$ denotes $T\sum_{\omega_k} \int_{-\pi}^{\pi} \! \int_{-\pi}^{\pi} {\rm d} {\bm k}/(2 \pi)^2$ 
with $k$, which is a notation that combines the wave vector ${\bm k}$ and the fermion Matsubara 
frequency ${\rm i}\omega_k$. The single-particle Green's function $G(k)$ is related to the self-energy $\Sigma(k)$ through the Dyson equation according to 
\begin{equation}
G(k) = [{\rm i} \omega_k + \mu -\varepsilon_{\bm k} - \Sigma(k)]^{-1},
\label{eq:Dyson}
\end{equation}
where $\varepsilon_{\bm k}$ denotes the noninteracting band dispersion.

For an explicit calculation of $\kappa$, 
we adopt the FLEX approximation~\cite{Bickers89,MMM03} without contributions of the particle-hole and particle-particle ladder diagrams, 
which is the simplest conserving approximation for $\Sigma$ that includes the nontrivial exchange-correlation effect~\cite{Baym62}. 
In this approximation, $\Sigma(k)$ is given by
\begin{equation}
\Sigma(k)=  -\frac{1}{2}\int_q G(k-q) \sum_{\alpha = {\rm c}, {\rm s}} V_{\alpha}(q),
\label{eq:eff_Sig}
\end{equation}
where $\int_{q}$ denotes $T\sum_{\omega_q} \int_{-\pi}^{\pi} \! \int_{-\pi}^{\pi} {\rm d} {\bm q}/(2 \pi)^2$
with $q$, which is a notation combining the wave vector ${\bm q}$ and the boson Matsubara 
frequency ${\rm i}\omega_q$. Here, $V_{\alpha}(q)$ represents the exchange-correlation interaction potential 
for $\alpha = {\rm c}$, ${\rm s}$, as
\begin{equation}
V_{\alpha}(q) = v_{\alpha}({\bm q}) / [ 1+v_{\alpha}({\bm q})\chi(q) ], 
\label{Eq:Fluctuation}
\end{equation}
with $v_{\rm c} ({\bm q})$ = $U+4V(\cos q_x + \cos q_y)$, 
$v_{\rm s} ({\bm q})$ = $-U$, and $\chi(q)$ = $-\int_{k} G(k+q)G(k)$. 
By combining eqs.~(\ref{eq:n})--(\ref{Eq:Fluctuation}), 
we arrive at a self-consistent determination of  $G$ for a given $n$ with a relative 
precision of $10^{-6}$. Hereinafter, we set $U=6$. 

\begin{figure}[t]
\begin{center}
\resizebox{55mm}{!}{\includegraphics{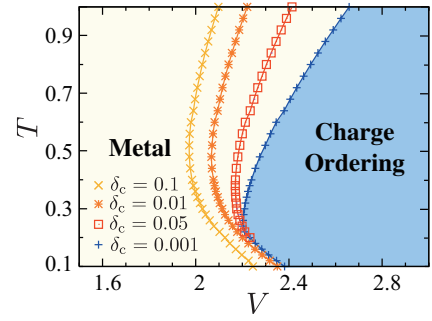}}
\end{center}
\vspace*{-1em}
\caption{(Color online)
A $T$-$V$ phase diagram for the EHM in the FLEX approximation with contour curves for $\delta_{\rm c}=0.001$, $0.01$, $0.05$, and 
$0.1$ (see legend for symbols) for $n = 3/2$ and $U = 6$. 
}
\label{fig:1}
\end{figure}

In the FLEX approximation, CO instability is determined by the divergence of $V_{\rm c}({\bm q}, 0)$. 
It is thus useful to define the ``distance" from the CO transition point as
\begin{equation}
\delta_{\rm c} \equiv 1+v_{\rm c}({\bm Q}_{\rm CO})\chi({\bm Q}_{\rm CO}, 0),
\label{delta}
\end{equation}
where ${\bm Q}_{\rm CO}$ is the wave vector specifying the CO pattern at which $|V_{\rm c}({\bm q}, 0)|$ is maximized. 
In Fig.~\ref{fig:1}, we show the contour curves of $\delta_{\rm c}$ on a $T$-$V$ phase diagram  
for $n = 3/2$, where the CO transition curve is determined by $\delta_{\rm c} =0$. 
Along the transition curve in Fig.~\ref{fig:1}, 
we consistently find instability to the unique pattern of checker-board-type CO specified by ${\bm Q}_{\rm CO} = (\pi, \pi)$ 
in accordance with previous studies.~\cite{ref:reentrant} 
Because ${\bm Q}_{\rm CO}$ is in disagreement with a maximum position of $\chi({\bm q},0)$ and 
is equated to the minimum position of $v_c({\bm q})$, the present CO instability is 
caused by $V$, which obliges electrons in nearest-neighbor sites to avoid each other in real space.~\cite{Seo04} 
This disagreement is a generic feature of CO and has been discussed for models with several different CO patterns 
in different contexts.\cite{Kuroki}.

From Fig.~\ref{fig:1}, we see that the intervals between the various $\delta_{\rm c}$ contour curves become much larger for 
a small $\delta_{\rm c}$ and a {\it nonzero} $T$; typically, $\delta_{\rm c}  \lesssim  0.05$ and $T \gtrsim  0.2$. 
Such increasing intervals are an essential feature of critical phenomena in itinerant electron systems
and originate from the self-consistent determination of $\delta_{\rm c}$, which includes a feedback effect for fluctuations~\cite{Moriya85}.
This self-consistency is partially taken into account through the determination of $G$ in the present FLEX approximation. 
Note that the region in which the interval enlargement occurs roughly corresponds to a quasiclassical regime governed by 
non-Fermi-liquid behavior in the theory of quantum critical phenomena~\cite{Hertz76}.  
In fact, for the EHM, the non-Fermi-liquid behavior and its related phenomena occur 
in regions dominated by quasiclassical CO fluctuations~\cite{Merino06}. 

\begin{figure}[t]
\begin{center}
\resizebox{55mm}{!}{\includegraphics{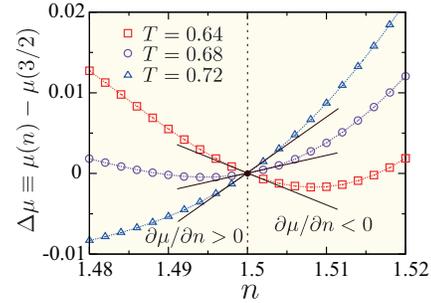}}
\end{center}
\vspace*{-1em}
\caption{(Color online) Chemical potential $\Delta \mu\equiv \mu(n)-\mu(3/2)$
as a function of  $n$ at 
$V=2.2$ for $T=0.72$ (triangles), 
$T=0.68$ (circles), and $T=0.64$ (squares). 
The tangent of each curve at $n = 3/2$ is shown by the solid lines.
}
\label{fig:2}
\end{figure}

\begin{figure}[t]
\begin{center}
\resizebox{55mm}{!}{\includegraphics{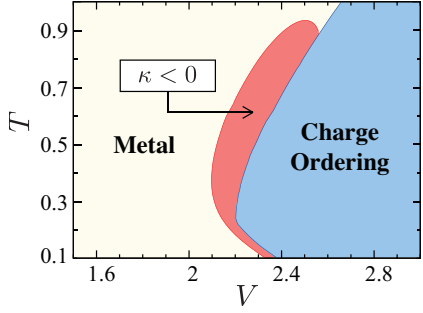}}
\end{center}
\vspace*{-1em}
\caption{(Color online) Region of $\kappa <0$ (shaded area indicated by arrow) on the same $T$-$V$ phase 
diagram as in Fig.~\ref{fig:1}.}
\label{fig:3}
\end{figure}

In Fig.~\ref{fig:2}, we plot $\Delta \mu \equiv \mu (n) - \mu(3/2)$  
as a function of $n$ for $T=0.72$, $T=0.68$, and $T=0.64$ with $V=2.2$.
For $T=0.72$, $\Delta \mu$ is a monotonic function with a positive slope
at $n = 3/2$. Upon decreasing the temperature to $T=0.68$, however, the slope 
at $n=3/2$ decreases, and at $T=0.64$, it is negative. The negative slope 
of the $\mu$-$n$ curve corresponds to a {\it negative} $\kappa$ because
$\kappa \equiv n^{-2} (\partial \mu / \partial n)^{-1} < 0$. In Fig.~\ref{fig:3}, 
we present the entire region of $\kappa <0$ on the same $T$-$V$ phase diagram 
as shown in Fig.~\ref{fig:1}, and we find that this region is in the quasiclassical regime 
mentioned in the last paragraph for quantum critical CO phenomena. 
Here, we emphasize that $\kappa$ becomes negative for a nonzero $T$ 
even though it is positive at $T = 0$.

To clarify the reason for $\kappa<0$, we recall that the interacting Fermi 
surface is determined by $\mu = \varepsilon_{{\bm k}_{\rm F}} + \Sigma ({\bm k}_{\rm F}, 0)$ 
at $T = 0$, where ${\bm k}_{\rm F}$ is the Fermi wave vector. A similar equation is available even for 
low but finite temperatures; $\mu$ is then approximated using
$\mu \approx \varepsilon_{{\bm k}_0} + {\rm Re} \Sigma^{\rm R} ({\bm k}_0, 0)$, where ${\bm k}_0 \equiv {\bm q}_0/2$ with ${\bm q}_0$ being the wave vector at which 
$\chi({\bm q}, 0)$ is maximized and $\Sigma^{\rm R} ({\bm k}_0, 0)$ is the analytic continuation of 
$\Sigma ({\bm k}_0, {\rm i} \omega_k)$ with $\omega_k > 0$ onto the origin of the complex frequency plane. 
Therefore, we can attribute the negative $\kappa$ to a rapid decrease in
${\rm Re} \Sigma^{\rm R} ({\bm k}_0, 0)$ with increasing $n$, which overcomes the increase in $\varepsilon_{{\bm k}_0}$. 

On the other hand, from eq.~(\ref{eq:eff_Sig}), we can see that, when $\delta_{\rm c} \to 0$ and $T \neq 0$, 
${\rm Re} \Sigma^{\rm R} ({\bm k}_0, 0)$ 
has a positive contribution in proportion to ${\rm Re} G^{\rm R}({\bm k}_0 - {\bm Q}_{\rm CO}, 0) \ln (1/\delta_{\rm c})$.
Therefore, we can attribute the rapid decrease in ${\rm Re} \Sigma^{\rm R} ({\bm k}_0, 0)$ to the increase in $\delta_{\rm c}$ with 
increasing $n$ in a quantum critical regime. This increase in $\delta_{\rm c}$ results from the combined effects of the 
intersite Coulomb repulsion, which makes the position of ${\bm Q}_{\rm CO}$ insensitive to variations in $n$, and the existence 
of the Fermi surface with a diameter smaller than $|{\bm Q}_{\rm CO}|$, as explained below. 

Because ${\bm Q}_{\rm CO} = (\pi, \pi)$ is not sensitive to slight variations in $n$ from $n = 3/2$ for which 
${\bm q}_0 = (\pi, \pi/2)$, $\delta_{\rm c}$ 
as defined by eq.~(\ref{delta}) 
depends on $n$ only through $\chi({\bm Q}_{\rm CO}, 0)$. 
Thus, because $|{\bm q}_0|$ corresponds to the diameter of the Fermi surface and decreases smaller as $n$ increases, 
we find that ${\bm q}_0$, i.e., the peak position of $\chi({\bm q}, 0)$, moves away from ${\bm Q}_{\rm CO}$ with increasing $n$. 
This variation in ${\bm q}_0$ reasonably reduces $\chi({\bm Q}_{\rm CO}, 0)$, so that $\delta_{\rm c}$ increases with $n$ 
into an off-critical regime, which leads to a negative $\kappa$. 
Since the discussion to this point does not depend on the details of the model,  
a negative $\kappa$ is expected in general for models that display a CO transition specified by a ${\bm Q}_{\rm CO}$ whose 
magnitude is large compared with the diameter of the Fermi surface. 

It is important to comment on $\kappa$ for the EHM 
in the limit of $V  = 0$, i.e., the Hubbard model (HM), in which spin-density-wave 
(SDW) fluctuations develop near the half filling ($n = 1$). 
By using the quantum Monte-Carlo method~\cite{FI92} and FLEX approximation~\cite{MMM03}, 
it has been shown that $\kappa$ is always {\it positive} for the 2D HM even though it tends to diverge for $n \to 1$. 
In the FLEX approximation, eq.~(\ref{eq:eff_Sig}) leads to the fact that 
${\rm Re} \Sigma^{\rm R} ({\bm k}_0, 0)$ has a contribution proportional to ${\rm Re} G^{\rm R}({\bm k}_0 - {\bm Q}_{\rm SDW}, 0) 
\ln (1/\delta_{\rm s})$ for $\delta_{\rm s} \to 0$ with $T \neq 0$, where $\delta_{\rm s}$ measures the ``distance" from the SDW quantum critical point.
It is, however, noted that the SDW wave vector ${\bm Q}_{\rm SDW}$ is the nesting vector of the Fermi surface, 
i.e., ${\bm Q}_{\rm SDW} = 2 {\bm k}_0$ and then ${\rm Re} G^{\rm R}({\bm k}_0 - {\bm Q}_{\rm SDW}, 0)  = {\rm Re} G^{\rm R}(- {\bm k}_0, 0) \approx 0$. 
Therefore, the singular contribution $\propto \ln (1/\delta_{\rm s})$ almost vanishes.  
This is the main reason why SDW fluctuations do not lead to a negative $\kappa$, 
while CO fluctuations do.

For the EHM defined by eq.~(\ref{eq:ham}), a negative $\kappa$ provides a short-range interaction to promote 
the spatial electron density fluctuations $\delta n({\bm r})$, which implies a macroscopic PS due to spinodal decomposition. 
This macroscopic PS is, however, cured because the short-range interaction competes with 
the long-range Coulomb interaction between $\delta n({\bm r})$ promoted by the short-range interaction. 
This type of arrested PS is called Coulomb frustrated PS. 

To see how our results for $\kappa$ within the EHM can lead to Coulomb frustrated PS,   
we first consider the following expansion of the free energy 
with respect to $\delta n({\bm r})$:
\begin{align}
\delta F = &
\frac{1}{2 D_{\rm F}}
\int d {\bm r}
\left[ \frac{\kappa_{\rm F}}{\kappa} \delta n({\bm r})^2 + 
l_0^2 |\nabla \delta n({\bm r})|^2  
\right] 
\nonumber \\
& + \frac{e^2}{2 \epsilon}\int {\rm d} {\bm r} {\rm d} {\bm r}'
\frac{\delta n({\bm r}) \delta n({\bm r'})}{|{\bm r}-{\bm r}'|},
\label{eq:GL-coordinate}
\end{align}
where $D_{\rm F}$ and $\kappa_{\rm F}$ denote the density of states 
and the compressibility for a noninteracting system, respectively. 
In eq.~(\ref{eq:GL-coordinate}), we consider the first and second terms to be the free-energy 
variation for the EHM due to the slowly varying spatial fluctuations.
The last term (with an electronic charge $e$ and a dielectric constant $\epsilon$) is taken to be the free energy due to long-range Coulomb interaction
that frustrates the PS and is absent in the EHM.
By using the Fourier transform 
$n ({\bm q}) = \int \delta n({\bm r}) {\rm e}^{- {\rm i} {\bm q} \cdot {\bm r} }$, 
eq.~(\ref{eq:GL-coordinate}) can be rewritten as     
\begin{align}
\delta F = 
\frac{1}{2 D_{\rm F}}
\sum_{{\bm q} \neq {\bm 0}}
\left[ 
\frac{\kappa_{\rm F}}{\kappa} + |{\bm q}|^2 l_0^2 + \frac{2}{|{\bm q}| a_{\rm B}^*}
\right] 
n({\bm q}) n(-{\bm q}),
\label{eq:GL-Fourier}
\end{align}
where the effective Bohr radius is $a_{\rm B}^* = \epsilon/(\pi e^2 D_{\rm F})$.  
The quantity in square brackets in eq.~(\ref{eq:GL-Fourier}) is minimized for 
$\kappa_{\rm F} / \kappa +3(l_0/a_{\rm B}^*)^{2/3}$ 
at $|{\bm q}| =  (a_{\rm B}^* l_0^2)^{-1/3} \neq 0$. 
Thus, the system becomes unstable 
against the spontaneous spatial modulation of $n({\bm q}) \neq 0$ with $|{\bm q}| =  (a_{\rm B}^* l_0^2)^{-1/3}$
for $\kappa_{\rm F}/ \kappa < - 3(l_0/a_{\rm B}^*)^{2/3}$ 
if there is no coupling between fluctuations associated with this phase transition.   
Note that our FLEX calculation of $\kappa_{\rm F} / \kappa$ for the EHM 
does not include any effect of these fluctuations since they are induced 
by the competition between the PS and long-range Coulomb interaction.

By adding the quartic term to eq.~(\ref{eq:GL-coordinate}), we can study the effects of the coupling between these fluctuations. 
Defining $\varphi({\bm r}) \equiv (l_0/\sqrt{D_{\rm F}}) \delta n({\bm r})$, we then consider the Hamiltonian 
with the coupling constant $u$ as
\begin{align}
{\cal H} = &
\frac{1}{2}
\int {\rm d} {\bm r}
\left[ r_0 \varphi({\bm r})^2 + 
|\nabla \varphi({\bm r})|^2 +\frac{u}{2} \varphi({\bm r})^4
\right] 
\nonumber \\ & 
+ \frac{Q}{4 \pi}\int {\rm d} {\bm r} {\rm d} {\bm r}'
\frac{\varphi({\bm r}) \varphi({\bm r'})}{|{\bm r}-{\bm r}'|},
\label{eq: phi4}
\end{align}
where $r_0=l_0^{-2} \kappa_{\rm F}/\kappa$ and $Q=2/(l_0^2 a_{\rm B}^*)$. 
In the mean-field approximation for eq.~(\ref{eq: phi4}), the Fourier transform of the correlation function 
${\cal G} ({\bm r} -{\bm r}') = T^{-1} \langle \varphi ({\bm r}) \varphi ({\bm r}') \rangle$ 
is given by
${\cal G}({\bm q})=\left( r + |{\bm q}|^2 + Q/|{\bm q}| \right)^{-1}$ 
where the parameter $r$ must be determined self-consistently by
\begin{equation}
r = r_0 + u T \int \frac{{\rm d}^2 q}{(2 \pi)^2} {\cal G}({\bm q}).
\label{eq: Hartree}
\end{equation}
Then ${\cal G}({\bm q})$ has a peak at approximately the modulation wave number $|{\bm q}| =q_{\rm m} \equiv \sqrt{-r/3}$ 
and can be approximated as     
\begin{equation}
{\cal G}({\bm q}) \simeq 
\frac{1}{|{\bm q}|} 
\left[
\frac{|{\bm q}|^2 - q_{\rm m}^2}{|{\bm q}|^3 - 3 q_{\rm m}^2|{\bm q}| + Q}  
+ \frac{q_{\rm m}/3}{(|{\bm q}|-q_{\rm m})^2+\xi^{-2}}
\right],
\label{eq:correlation2}
\end{equation}
where the inverse correlation length $\xi^{-1}$ is given by
\begin{equation}
\xi^{-1} 
=l_0^{-1} \sqrt{
2 \left[ (l_0/a_{\rm B}^*) - (q_{\rm m}l_0)^3 \right]
/(3 q_{\rm m} l_ 0)}. 
\label{eq: xi1}
\end{equation}
Substituting eq.~(\ref{eq:correlation2}) into eq.~(\ref{eq: Hartree}), 
we can obtain $\xi^{-1}$ in the form 
\begin{equation}
\xi^{-1} = q_{\rm m} 
\frac{\displaystyle{ \frac{A}{2 \pi} \frac{T}{t}}
}{
\displaystyle{ \left| \frac{\kappa_{\rm F}}{\kappa} \right| 
- 3 (q_{\rm m} l_0)^2 - \frac{3 A}{2 \pi^2} \frac{T}{t} 
\ln \left[ \frac{\Lambda l_0}{ (2 l_0 / a_{\rm B}^*)^{1/3} } \right]}
}, 
\label{eq: xi2}
\end{equation}
where $A = (\pi/3) u l_0^2 t$ and $\Lambda$ is the cutoff wave number. 
Because the right-hand sides of eqs.~(\ref{eq: xi1}) and (\ref{eq: xi2}) 
are equal to each other, we have an equation for determining $q_{\rm m} = \sqrt{-r/3}$, 
and then the period of modulation is given by $l_{\rm m} = 2\pi / q_{\rm m}$. 
Using the solution of this equation for eq.~(\ref{eq: xi2}), 
we can obtain the correlation length $\xi$. Note that $\xi$ thus obtained 
never diverges unless $T = 0$; namely, the transition 
into the modulated phase does not occur even though $\kappa_{\rm F}/ \kappa < - 3(l_0/a_{\rm B}^*)^{2/3}$ 
owing to the feedback effect for isotropic fluctuations 
peaked at the modulation wave number $q_{\rm m} \neq 0$.\cite{Nussinov}

\begin{figure}[lt]
\begin{center}
\resizebox{55mm}{!}{\includegraphics{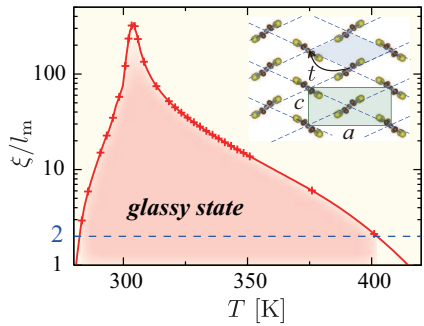}}
\end{center}
\vspace*{-1em}
\caption{(Color online) $\xi/l_{\rm m}$ versus $T$ for $V/t=2.2$ and $U/t=6$ with $t=0.108$~{\rm eV}  
assuming that $A = 1$ and $l_0/a_{\rm B}^* = 1$ ($a_{\rm B}^*=2.56\AA$). 
The inset shows the 2D arrangement of the ET molecules with $t$, $a$, and $c$ indicated.}
\label{fig:4}
\end{figure}

In this situation, Schmalian {\it et al}. have shown that 
glassiness emerges instead of the spontaneous density modulation 
if the ratio $\xi/l_{\rm m}$ is larger than the critical value close to $2$.~\cite{Schmalian00, Westfahl01} 
This emergence of a glassy state is signaled by the occurrence of an exponentially large number of metastable states and 
long-time correlations characterized by the correlation function 
${\cal F} ({\bm r} -{\bm r}') = T^{-1}\lim_{t \to \infty} \langle \varphi ({\bm r},t) \varphi ({\bm r}',0) \rangle 
\neq 0$, leading to an extremely slow relaxation and an intermedeate length scale of $\xi$. 
Therefore, we can demonstrate that our system has glassy dynamics by showing that $\xi/l_{\rm m} >2$.  

By using our numerical result of $\kappa_{\rm F}/\kappa$ for the $3/4$-filled EHM with $V/t = 2.2$ and $U/t = 6$, 
we can actually evaluate $\xi/l_{\rm m}$ for quasi-2D organic conductors on the verge of CO 
(with the recovery of $a_{\rm L}$ and $t$).
For $\theta$-ET$_2$CsZn(SCN)$_4$, the transfer integral and lattice constants shown in the inset of Fig.~\ref{fig:4} 
are $t$ = $0.108~{\rm eV}$, $a$ = $9.82~\mbox{\AA}$, and $c$ = $4.87~\mbox{\AA}$~\cite{Mori98}.
We then obtain $\epsilon = 11$ from $V=e^2/(\epsilon r_{\rm NN}) = 2.2 t$ and $D_{\rm F} = 0.247/(t a_{\rm L}^2 )$ 
for the square lattice ($r_{\rm NN}=\sqrt{a^2+c^2}$ and $a_{\rm L}= \sqrt{ac/2}$).  
The effective Bohr radius is then evaluated as $a_{\rm B}^*=\epsilon/(\pi e^2 D_{\rm F})=2.56\AA$. 
The cutoff wave number $\Lambda$ should be comparable to the magnitude of the reciprocal lattice vector 
and we choose $\Lambda = 2 \sqrt{2} \pi/a_{\rm L}$. 
Numerical values of $A$ and $l_0/a_{\rm B}^*$ are expected to be on the order of $1$. 
Here, we take $A = 1$ and $l_0/a_{\rm B}^*  = 1$ as a rough estimation of $\xi/l_{\rm m}$. 
By inserting these values together with our results for $\kappa_{\rm F}/\kappa$ into eq.~(\ref{eq: xi2}), 
we obtain $\xi/l_{\rm m}$ as a function of $T$, which is shown in Fig. \ref{fig:4}. 
From this figure and Fig. \ref{fig:1}, we see that $\xi/l_{\rm m}$ has a peak at approximately $T=320$~[K] at which 
$\delta_{\rm c}$ takes the smallest value. We find that $\xi/l_{\rm m}$ is larger than $2$, i.e., the value of criterion 
for glassiness, for a broad range of $T$ around this temperature. 
Note that in our calculation, $l_{\rm m}$ is on the order of nanometers, 
so that $\xi$ can be over hundreds of nanometers, which seems to be too large for the experimentally 
observed length scale of the intermediate-range order.

In actual quasi-2D organic conductors, because of the screening by electrons in other layers, 
the intralayer Coulomb interaction is reduced to a screened Coulomb interaction with 
the screening length $l_{\rm s}$ nearly equal to the interlayer distance $b$.\cite{Nakamura}
Therefore, the above values of $\xi/l_{\rm m}$ may be overestimated. 
It is, however, noted that the formation of a glassy state is virtually unchanged, 
even considering the screened Coulomb interaction instead of the long-range Coulomb interaction, 
provided that the condition $(l_0^2 a_{\rm B}^*)^{1/3} \ll l_{\rm s}$ is satisfied.~\cite{Rev_Schmalian} 
This condition is, in fact, satisfied for the $\theta$-ET organic conductors; 
for example, $(l_0^2 a_{\rm B}^*)^{1/3} \sim a_{\rm B}^* =2.56 \AA$ is 
much smaller than $l_{\rm s} \sim b = 43.4 \AA$~\cite{Mori98} for $\theta$-ET$_2$CsZn(SCN)$_4$.
Hence, our mechanism can produce a glassy state for actual $\theta$-ET organic conductors on the verge of CO. 

%acknowledge
The authors thank T. Kato, H. Mori, H. Seo, and H. Shinaoka for helpful discussions. 
This work was supported by a Grant-in-Aid for Scientific Research 
in the Priority Area of Molecular Conductors (No. 20110003, 20110004, 21110510) from 
the Ministry of Education, Culture, Sports, Science and Technology.

%%%%%%%%%%%%%%%%%%%%%%%%%%%%%%%%%< References >%%%%%%%%%%%%%%%%%%%%%%%%%%%%%%%%%


\begin{thebibliography}{99}
% Rreviews
% Rreviews
\bibitem{Seo04} For reviews, see H. Seo {\it et al.}: Chem. Rev. {\bf 104} (2004) 5005; 
H. Seo {\it et al.}: J. Phys. Soc. Jpn. {\bf 75} (2006) 051009.
\bibitem{Singleton02} J. Singleton and C. Mielke: Contemp. Phys. {\bf 43} (2002) 63.
\bibitem{Sasaki} The effect of artificial disorder has been studied in K. Sano {\it et al.}; Phys. Rev. Lett. {\bf 104} (2010) 217003.

%Expriments on an extremely slow relaxation
\bibitem{Miyagawa00} K. Miyagawa {\it et al.}: Phys. Rev. B {\bf 62} (2000) R7679; 
R. Chiba {\it et al.}: Phys. Rev. Lett. {\bf 93} (2004) 216405; R. Chiba {\it et al.}: Phys. Rev. B {\bf 77} (2008) 115113.

%Expriments on the size of domains
\bibitem{Watanabe99} M. Watanabe {\it et al.}: J. Phys. Soc. Jpn. {\bf 68} (1999) 2654; 
M. Watanabe {\it et al.}: Synth. Met. {\bf 135-136} (2003) 665; M. Watanabe {\it et al.}: J. Phys. Soc. Jpn. {\bf 74} (2005) 2011.
% Thyristor in theta
\bibitem{Sawano05} F. Sawano {\it et al.}: Nature {\bf 437} (2005) 522; 
Y. Nogami {\it et al.}: J. Phys. Soc. Jpn. {\bf 79} (2010) 044606.

% ED
\bibitem{Merino05} J. Merino {\it et al.}: Phys. Rev. B {\bf 71} (2005) 125111;
% VMC
H. Watanabe and M. Ogata: J. Phys. Soc. Jpn. {\bf 75} (2006) 063702;
% DMRG, ED
S. Nishimoto {\it et al.}: Phys. Rev. B {\bf 78} (2008) 035113.

%Theory on a glassy state induced by frustrated interaction
\bibitem{Schmalian00} J. Schmalian and P. G. Wolynes: Phys. Rev. Lett. {\bf 85} (2000) 836.
\bibitem{Westfahl01} H. Westfahl, Jr. {\it et al.}: Phys. Rev. B {\bf 64} (2001) 174203.
% phase separation for models of strongly-correlated electrons
\bibitem{Emery90} V. J. Emery {\it et al.}: Phys. Rev. Lett. {\bf 64} (1990) 475; 
H. Yokoyama and M. Ogata: J. Phys. Soc. Jpn. {\bf 65} (1996) 3615. 
\bibitem{Yunoki98} S. Yunoki {\it et al.}: Phys. Rev. Lett.  {\bf 80} (1998) 845. 
\bibitem{Vosko80} S. H. Vosko {\it et al.}: Can. J. Phys. {\bf 58} (1980) 1200; 
Y. Takada: Phys. Rev. B {\bf 43} (1991) 5979.

% frustrated phase separation scenario
\bibitem{Ortix08} C. Ortix {\it et al.}: Phys. Rev. Lett. {\bf 100} (2008) 246402.
\bibitem{Emery93} V. J. Emery and S. A. Kivelson: Physica C {\bf 209} (1993) 597.
\bibitem{Yunoki99} A. Moreo {\it et al.}: Science {\bf 283} (1999) 2034.
\bibitem{Maebashi09} H. Maebashi and Y. Takada: J. Phys.: Condens. Matter {\bf 21} (2009) 064205; H. Maebashi and Y. Takada: J. Phys. Soc. Jpn. {\bf 78} (2009) 053706.

%FLEX approximation
\bibitem{Bickers89} N. E. Bickers {\it et al.}: Phys. Rev. Lett. {\bf 62} (1989) 961. 
\bibitem{MMM03}
K. Morita {\it et al.}: J. Phys. Soc. Jpn. {\bf 72} (2003) 3164. 
\bibitem{Baym62} G. Baym: Phys. Rev. {\bf 127} (1962) 1391.
\bibitem{ref:reentrant} J. Merino and R. H. McKenzie: Phys. Rev. Lett. {\bf 87} (2001) 237002; A. Kobayashi {\it et al.}: J. Phys. Soc. Jpn. {\bf 73} (2004) 1115.
\bibitem{Kuroki} K. Kuroki: J. Phys. Soc. Jpn. {\bf 75} (2006) 114716; K. Yoshimi {\it et al.}: J. Phys. Soc. Jpn. {\bf 80} (2011) 123707.

%quantum critical phenomena	
\bibitem{Moriya85} T. Moriya: {\it Spin Fluctuations in Itinerant Electron Magnetism} (Springer-Verlag, Berlin, 1985).
\bibitem{Hertz76} J. A. Hertz: Phys. Rev. B {\bf 14} (1976) 1165; 
A. J. Millis: Phys. Rev. B {\bf 48} (1993) 7183.

%critical phenomena for EHM
\bibitem{Merino06} J. Merino {\it et al.}: Phys. Rev. Lett. {\bf 96} (2006) 216402; 
K. Yoshimi {\it et al.}: J. Phys. Soc. Jpn. {\bf 78} (2009) 104002;
L. Cano-Cort\'es {\it et al.}: Phys. Rev. Lett. {\bf 105} (2010) 036405.

\bibitem{FI92} N. Furukawa and M. Imada: J. Phys. Soc. Jpn. {\bf 61}
 (1992) 3331; N. Furukawa and M. Imada: J. Phys. Soc. Jpn. {\bf 62} (1993) 2557.

\bibitem{Nussinov} For more detail, see G. Tarjus {\it et al.}: 
J. Phys.: Condens. Matter {\bf 17} (2005) R1143.

% long-range Coulomb interaction in organic conductors
\bibitem{Mori98} H. Mori {\it et al.}: Phys. Rev. B {\bf 57} (1998) 12023.

\bibitem{Nakamura} H. Shinaoka {\it et al.}: J. Phys. Soc. Jpn. {\bf 81} (2012) 034701.

\bibitem{Rev_Schmalian} M. Dzero {\it et al.}: {\it Structural Glasses and Supercooled Liquids: Theory, Experiment, and Applications}, ed. P. G. Wolynes and V. Lubchenko (Wiley, Hoboken, 2012) Chap. 5.



\end{thebibliography}
\end{document}